\documentclass[
 reprint,
 superscriptaddress,
 amsmath,
 amssymb,
 aps,
 pra,
]{revtex4-2}

\usepackage[dvipdfmx]{graphicx}
\usepackage{bm}
\usepackage{color}
\usepackage{amssymb}
\usepackage{mathtools}
\usepackage{empheq}
\usepackage{braket}
\usepackage{ulem} 
\renewcommand{\sout}{\bgroup\markoverwith{\textcolor{red}{\rule[0.5ex]{2pt}{1pt}}}\ULon}

\begin{document}

\preprint{APS/123-QED}

\title{Speed Limit of Efficient Cavity-Mediated Adiabatic Transfer}

\author{Akinori Suenaga}
\email{aki.physic.al@gmail.com}
\affiliation{Department of Applied Physics, Waseda University, Okubo, Shinjuku, Tokyo 169-8555, Japan}
\author{Takeru Utsugi}
\email{utsugitakeru@akane.waseda.jp}
\affiliation{Department of Applied Physics, Waseda University, Okubo, Shinjuku, Tokyo 169-8555, Japan}
\author{Rui Asaoka}
\affiliation{Computer and Data Science Laboratories, NTT Corporation, Musashino 180-8585, Japan}
\author{Yuuki Tokunaga}
\affiliation{Computer and Data Science Laboratories, NTT Corporation, Musashino 180-8585, Japan}
\author{Rina Kanamoto}
\affiliation{Department of Physics, Meiji University, Kawasaki, Kanagawa 214-8571, Japan}
\author{Takao Aoki}
\affiliation{Department of Applied Physics, Waseda University, Okubo, Shinjuku, Tokyo 169-8555, Japan}

\date{\today}

\begin{abstract}
    Cavity-mediated adiabatic transfer (CMAT) is a robust way to perform a two-qubit gate between trapped atoms inside an optical cavity.
    In the previous study by Goto and Ichimura [H. Goto and K. Ichimura, \href{https://journals.aps.org/pra/abstract/10.1103/PhysRevA.77.013816}{Phys. Rev. A {\bf 77}, 013816 (2008).}], the upper bound of success probability of CMAT was shown where the operation is adiabatically slow.
    For practical applications, however, it is crucial to operate CMAT as fast as possible without sacrificing the success probability.
    In this paper, we investigate the operational speed limit of CMAT conditioned on the success probability being close to the upper bound.
    In CMAT both the adiabatic condition and the decay of atoms and cavity modes limit the operational speed.
    We show which of these two conditions more severely limits the operational speed in each cavity-QED parameter region, and find that the maximal operational speed, which is proportional to $\gamma\sqrt{C}$, is achieved when the influence of cavity decay is dominant compared to spontaneous emission, where $\gamma$ and $C$ are spontaneous emission rate and cooperativity.
\end{abstract}

\pacs{Valid PACS appear here}
\maketitle

\section{Introduction\label{sec_Int}}
\begin{figure}[t]
        \centering
        \includegraphics[clip,width=6cm]{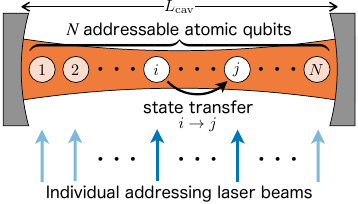}
        \caption{Schematic of CQED-based quantum computing with individually addressable atoms, with cavity length $L_\mathrm{cav}$.
        Each atom in the cavity represents a single qubit. 
        The individual atomic qubits are controlled by focused laser beams.
        CMAT enables two-qubit gates between atoms via cavity photon.
        }
        \label{Many-atomCavityQED}
    \end{figure}

    Cavity-mediated adiabatic transfer (CMAT) provides a robust way to transfer quantum states between atoms~\cite{Pellizzari1995}.
    CMAT is the process of adiabatically performing state transfer between atoms inside a cavity via a cavity photon.
    The most promising application of CMAT is to enable two-qubit gates between individually addressable atoms~\cite{Pellizzari1995, You2003, Goto2004, Sangouard2005, Sangouard2006, Lacour2006, Lambropoulos2007, Ramette2022}, which can realize quantum computing based on cavity quantum electrodynamics (CQED), such as that shown in Fig.~\ref{Many-atomCavityQED}.
    A number of two-qubit gates utilizing CMAT have already been proposed, including controlled NOT gates~\cite{Pellizzari1995, Sangouard2006, Lambropoulos2007, Ramette2022}, controlled phase gates~\cite{You2003, Goto2004}, SWAP gates~\cite{Sangouard2005} and arbitrary-controlled unitary gates~\cite{Pellizzari1995, Goto2004, Lacour2006}.
    
    Quantum computing based on these gates has an advantage in qubit connectivity since any atoms in the cavity can interact with each other through the cavity photon~\cite{Ramette2022}.
    In this way, the scheme in Fig.~\ref{Many-atomCavityQED} is superior to other types of CQED-based quantum computing using $\pi$-phase flip reflection~\cite{Duan-Kimble2004,Duan-Kimble2005}, which have been recently demonstrated in experiment~\cite{Rempe2016,Rempe2021}.
    Moreover, CQED-based quantum computing is well known as an efficient platform for expanding the scale of computation by networking~\cite{Kimble2008, Ritter2012, Rempe2015, Rempe2021}.
    Thus, quantum computing using CMAT is also one of the prominent approaches to build a scalable platform.

    In practice, several error sources may prevent the success of CMAT. In a previous study, Goto and Ichimura have shown the upper bound of success probability of CMAT, which is characterized by the cooperativity $C$ as $\exp{(-2/\sqrt{C})}$~\cite{Goto2008}. Approaching this upper bound in CMAT is desirable since quantum computing requires a high success probability. However, there are additional requirements for practical quantum computing, such as operational speed. Thus, it is essential to elucidate the conditions of the control operation and the CQED parameters, ideally ensuring they meet multiple requirements simultaneously.

    In this paper, we explore the operational speed limit of CMAT conditioned on the success probability being close to the upper bound. If the CQED system is isolated, a sufficiently slow operational speed satisfying the adiabatic condition~\cite{MessiahBook1978,Bergmann1998} is the best way to achieve a high success probability. In practice, however, the duration time of a control laser in CMAT should be as short as possible because the CMAT operation must be performed in a sufficiently short span compared to the finite trap lifetime or the decoherence time of trapped atoms~\cite{McKeever2003,Rempe2009,Rempe2010,Goban2012,Kato2015,Rauschenbeutel2018,Nayak2019}.
    Moreover, photon loss through spontaneous emission and cavity decay also limit the operational speed since the time evolution of the populations of the atomic excited state and the cavity photon depend on the operational speed~\cite{Shahriar2001,Goto2008,Steane2000,Sangouard2005,Deng2007, Sangouard2006,Lacour2006}. For instance, the total photon loss can be minimized by balancing the above two effects~\cite{Molmer2003, Goto2008, Ramette2022}, which is achieved when the parameters of the control laser, namely the duration time $T$ and the intensity, satisfy $T = \alpha/{\Omega_0}^2$~\cite{Goto2008}, where $\alpha$ and $\Omega_0$ are the coefficient defined by CQED parameters and the Rabi frequency corresponding to the maximum intensity of the laser during the operation, respectively.
    Here we focus on two error sources, adiabatic condition violation~\cite{MessiahBook1978,Bergmann1998} and photon loss, and determine which more severely limits the operational speed in each CQED parameter region. We also investigate the maximal operational speed for the entire parameter region.

    This paper is organized as follows: 
    Section~\ref{sec_CMAT} explains the basic idea of CMAT, influence of photon loss, and photon-loss balancing condition.
    In Sec.~\ref{sec_Main}, we formulate the factors limiting the operational speed of CMAT.
    We also confirm the validity of the formulation by numerical simulations.
    Our conclusion is summarized in Sec.~\ref{sec_Conclusion}.

\section{Cavity-Mediated Adiabatic Transfer\label{sec_CMAT}}
    \subsection{Basic principle\label{sec_basic-principle}}
    \begin{figure}[t]
        \centering
        \includegraphics[clip,width=6cm]{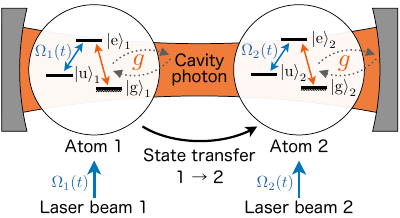}
        \caption{Conceptual diagram of CMAT.
        $\Ket{\mathrm{e}}_i$, $\Ket{\mathrm{g}}_i$, and $\Ket{\mathrm{u}}_i$ denote the excited, ground, and uncoupled states, respectively ($i=1,2$). 
            The $\Ket{\mathrm{e}}_i\leftrightarrow\Ket{\mathrm{g}}_i$ and $\Ket{\mathrm{e}}_i\leftrightarrow\Ket{\mathrm{u}}_i$ transitions are allowed, but the $\Ket{\mathrm{g}}_i\leftrightarrow\Ket{\mathrm{u}}_i$ transition is prohibited. The $\Ket{\mathrm{e}}_i\leftrightarrow\Ket{\mathrm{g}}_i$ transition is coupled to the cavity photon mode with the coupling constant $g$, while the $\Ket{\mathrm{e}}_i\leftrightarrow\Ket{\mathrm{u}}_i$ transition is driven by an external laser beam (classical field) with Rabi frequency $\Omega_i(t)$.
        }
        \label{CMAT_ConceptualDiagram}
    \end{figure}
    
    We first explain the idea of CMAT based on Refs.~\cite{Pellizzari1995, Lambropoulos2007}, focusing on the state transfer between two atoms trapped inside a cavity for simplicity.
    Figure~\ref{CMAT_ConceptualDiagram} shows a conceptual diagram of CMAT.
    The atoms are assumed to be identical and have a three-level internal structure.
    Each atomic state $\Ket{\mathrm{e}}_i$, $\Ket{\mathrm{g}}_i$, and $\Ket{\mathrm{u}}_i$ denotes excited, ground and uncoupled state, respectively ($i=1,2$).
    The qubit state of atom $i$ is represented by $\Ket{\mathrm{g}}_i$ and $\Ket{\mathrm{u}}_i$.
    Then the state transfer from atom 1 to atom 2 is expressed as
    \begin{eqnarray}
        &&\Bigl(
            C_\mathrm{g} \Ket{\mathrm{g}}_1 + C_\mathrm{u}\Ket{\mathrm{u}}_1
        \Bigr)\Ket{\mathrm{g}}_2 \Ket{0}_\mathrm{c}
        \nonumber\\
        &&\ \ \longrightarrow\ \ \ 
        \Ket{\mathrm{g}}_1\Bigl(
            C_\mathrm{g} \Ket{\mathrm{g}}_2 + C_\mathrm{u}\Ket{\mathrm{u}}_2
        \Bigr) \Ket{0}_\mathrm{c}
        ,
    \label{stateTransfer}
    \end{eqnarray}
    where $C_\mathrm{g}$ and $C_\mathrm{u}$ are arbitrary complex amplitudes that satisfy the normalization condition, and the third ket vector is the number state of the cavity photon.
    In CMAT, this state transfer is performed by utilizing the adiabatic passage of a dark state.
    
    The dark state is obtained when the $\Ket{\mathrm{e}}_i\leftrightarrow\Ket{\mathrm{g}}_i$ transition is coupled to the cavity photon mode with the coupling constant $g$, and the $\Ket{\mathrm{e}}_i\leftrightarrow\Ket{\mathrm{u}}_i$ transition is driven by an external control laser beam (classical field) with Rabi frequency $\Omega_i(t)$, as shown in Fig.~\ref{CMAT_ConceptualDiagram}.
    Note that here $\Omega_i(t)$ is real number for simplicity.
    In the frame rotating with the frequencies of the cavity and driving beams, the Hamiltonian of this system is given by
    \begin{eqnarray}
        \mathcal{H}(t) / \hbar &=&
        \mathrm{i} g \sum_{i=1,2} \Bigl(
            a\Ket{\mathrm{e}} {}_i {}_i {\Bra{\mathrm{g}}} - a^\dag\Ket{\mathrm{g}} {}_i {}_i \Bra{\mathrm{e}}
        \Bigr)
        \nonumber
        \\
        &&
        +\ \mathrm{i} \sum_{i=1,2} \Omega_i(t) \Bigl(
            \Ket{\mathrm{e}} {}_i {}_i \Bra{\mathrm{u}} - \Ket{\mathrm{u}} {}_i {}_i \Bra{\mathrm{e}}
        \Bigr)
        ,
        \label{Hamiltonian}
    \end{eqnarray}
    where the rotating-wave approximation is applied, and $a$ represents the annihilation operator of the cavity mode.
    Note that the influence of detuning is not taken into account in this model.
    \\\\
    
    The Hamiltonian in Eq.~\eqref{Hamiltonian} has an eigenstate with zero eigenvalue~\cite{Pellizzari1995}
    \begin{widetext}
    \begin{equation}
        \ket{\mathrm{D}(t)} = 
        \frac{g\Omega_2(t)\Ket{\mathrm{u}}_1\Ket{\mathrm{g}}_2\Ket{0}_\mathrm{c}
                + g\Omega_1(t)\Ket{\mathrm{g}}_1\Ket{\mathrm{u}}_2\Ket{0}_\mathrm{c}
                - \Omega_1(t)\Omega_2(t)\Ket{\mathrm{g}}_1\Ket{\mathrm{g}}_2\Ket{1}_\mathrm{c}
            }{\sqrt{ g^2{\Omega_1(t)}^2 + g^2{\Omega_2(t)}^2 + {\Omega_1(t)}^2 {\Omega_2(t)}^2 }
            }   ,
        \label{Eq_darkState}
    \end{equation}
    \end{widetext} 
    which corresponds to the dark state represented only by ground states $\Ket{\mathrm{g}}_1$ and $\Ket{\mathrm{u}}_1$.
    Note that the other eigenstates, when the total number of excitation is one, are given in Appendix~\ref{appendix_eigen_derivation}.
    According to the adiabatic theorem~\cite{MessiahBook1978}, the initially prepared state $\Ket{\mathrm{u}}_1\Ket{\mathrm{g}}_2\Ket{0}_\mathrm{c}$ transitions to $\Ket{\mathrm{g}}_1\Ket{\mathrm{u}}_2\Ket{0}_\mathrm{c}$ without deviating from the dark state in Eq.~\eqref{Eq_darkState} by adiabatically varying the Rabi frequencies such that
    \begin{equation}
        \left.\frac{\Omega_1}{\Omega_2}\right|_\text{initial}=0
        \ \ \ \to\ \ \ 
        \left.\frac{\Omega_1}{\Omega_2}\right|_\text{final}=\infty
        .
        \label{CMAToperation}
    \end{equation}
    As this operation does not change the state $\Ket{\mathrm{g}}_1\Ket{\mathrm{g}}_2\Ket{0}_\mathrm{c}$, the state transfer in Eq.~\eqref{stateTransfer} can be performed through this operation.

    \subsection{Formulation including photon loss\label{sec_photon-loss-error}}
    In practice, photon loss associated with spontaneous emission and cavity decay are inevitable error factors in CMAT.
    Here, we define the success probability of CMAT under the consideration of photon loss, based on a method using an effective Hamiltonian~\cite{Goto2008,Plenio1998,CarmichaelBook2008}.
    In CMAT, spontaneous emission and cavity decay are characterized by an effective Hamiltonian such that
    \begin{equation}
        \mathcal{H}_\mathrm{eff} = \mathcal{H}-\mathrm{i}\hbar V,\ \ 
        V = \gamma\Ket{\mathrm{e}}_{11}\langle\mathrm{e}|
        + \gamma\Ket{\mathrm{e}}_{22} \langle\mathrm{e}|
        + \kappa a^\dag a,
        \label{Eq_effectiveHamiltonian}
    \end{equation}
    where $\kappa$ and $\gamma$ are the amplitude decay rate of the cavity photon and the polarization decay rate of spontaneous emission, respectively.
    Note that, in the previous study of Ref.~\cite{Goto2008}, $\gamma$ differs by a factor of 2 due to distinct definition.
    The time evolution of the system under photon loss is obtained by solving the following Schr\"{o}dinger-like equation
    \begin{equation}
        \mathrm{i}\hbar\frac{d\ }{dt}\ket{\widetilde{\Psi}(t)}
        = \mathcal{H}_\mathrm{eff}(t)\ket{\widetilde{\Psi}(t)},
        \label{Eq_TimeEvolutionEquation}
    \end{equation}
    with the initial state $\Ket{\mathrm{u}}_1\Ket{\mathrm{g}}_2\Ket{0}_\mathrm{c}$.
    Note that $\ket{\widetilde{\Psi}(t)}$ is not normalized since $\mathcal{H}_\mathrm{eff}(t)$ is a non-Hermitian operator, and the normalized system state is obtained by
    \begin{equation}
        \Ket{\Psi(t)} = \frac{\ket{\widetilde{\Psi}(t)}}{\sqrt{\braket{\widetilde{\Psi}(t)|\widetilde{\Psi}(t)}}}.
    \end{equation}
    The probability of no photon loss is given by $\braket{\widetilde{\Psi}(t)|\widetilde{\Psi}(t)}$. 
    Thus the probability to obtain the ideal final state  $\Ket{\mathrm{g}}_1\Ket{\mathrm{u}}_2\Ket{0}_\mathrm{c}$ without photon loss, which is called success probability here, is defined as~\cite{Goto2008}
    \begin{equation}
        P_\mathrm{s} = \braket{\widetilde{\Psi}(t=\infty)|\widetilde{\Psi}(t=\infty)}
        \left|\braket{\Psi(t=\infty)|\mathrm{g}}_1\Ket{\mathrm{u}}_2\Ket{0}_\mathrm{c}\right|^2,
        \label{Eq_success_probability}
    \end{equation}
    where $\left|\braket{\Psi(t=\infty)|\mathrm{g}}_1\Ket{\mathrm{u}}_2\Ket{0}_\mathrm{c}\right|^2$ represents the fidelity to the ideal final state.

    \subsection{Photon-loss balancing\label{sec_photon-loss-blancing}}
    Photon loss due to spontaneous emission and cavity decay occurs via finite populations in the excited state and cavity field.
    In CMAT, atomic excitation is induced by the violation of the adiabatic condition.
    On the other hand, the population of cavity field $p_\mathrm{a}$ is given by 
    \begin{equation}
        p_\mathrm{a} = \frac{{\Omega_1(t)}^2{\Omega_2(t)}^2}{g^2{\Omega_1(t)}^2+g^2{\Omega_2(t)}^2+{\Omega_1(t)}^2{\Omega_2(t)}^2},
        \label{Eq_Cavity-photon-porbability}
    \end{equation}
    from the coefficient of $\Ket{\mathrm{g}}_1\Ket{\mathrm{g}}_2\Ket{1}_\mathrm{c}$ in the dark state.
    This indicates that the photon population can be suppressed for $\Omega_0/g\ll 1$~\cite{Goto2008, Steane2000, Sangouard2005, Deng2007, Sangouard2006}, where $\Omega_0$ is the Rabi frequency corresponding to the maximum laser intensity.
    Thus, it is preferable to both satisfy the adiabatic condition and suppress the photon population for reducing the total photon loss.
    
    In Ref.~\cite{Goto2008}, under the approximation that both the adiabatic condition and $\Omega_0/g\ll 1$ are well satisfied, the photon loss probability $P_\mathrm{pl}$ is derived as
    \begin{eqnarray}
        P_\mathrm{pl} = 1-e^{-\beta},\ \ \ \ \beta =
        \kappa\cdot\frac{\tau{\Omega_0}^2}{g^2} + \gamma \cdot \frac{2}{\tau{\Omega_0}^2},
        \label{Eq_beta}
    \end{eqnarray}
    where the laser is controlled following reasonable functions for high fidelity~\cite{Goto2008}:
    \begin{equation}
        \Omega_i(t)=\Omega_0f_i(t)\ \ (i=1,2),\ \ \ \ 
        \begin{cases}
        f_1(t) = \displaystyle\frac{e^{t/\tau}}{\sqrt{e^{2t/\tau}+1}}\\
        f_2(t) = \displaystyle\frac{1}{\sqrt{e^{2t/\tau}+1}}.
        \end{cases}
        \label{Eq_concrete_f1f2}
    \end{equation}
    Increasing $\tau$ in Eq.~\eqref{Eq_concrete_f1f2} means increasing the total duration time of the control laser.
    The first term and the second term of $\beta$ in Eq.~\eqref{Eq_beta} stem from cavity decay and spontaneous emission, respectively.
    It was shown in Ref.~\cite{Goto2008} that the photon loss probability $P_\mathrm{pl}$ is minimized when these two terms are equal satisfying the photon-loss balancing condition,
    \begin{equation}
        \tau{\Omega_0}^2 = g \sqrt{\frac{2\gamma}{\kappa}},
        \label{Eq_PL-balancing-condition}
    \end{equation}
    and the minimal value of $P_\mathrm{pl}$ is $1-\exp{(-2/\sqrt{C})}$, where the cooperativity is defined by $C=g^2/(2\kappa\gamma)$\cite{Rempe2015}.
    That is, the upper bound of the success probability is $\exp{(-2/\sqrt{C})}$.
    
\section{Operational Speed Limit in CMAT\label{sec_Main}}
    \subsection{Formulation \label{sec_formulation-operation-speed-restriction}}
    Here we investigate the operational speed limit of CMAT conditioned on the success probability being close to the upper bound.
    While the adiabatic condition obviously restricts the operational speed of CMAT, it is not clear how the photon loss does.
    Our main focus here is which of the error sources, the violation of adiabatic condition and photon loss, more severely limits the operational speed in each CQED parameter region.

    A typical representation of the adiabatic condition is given by
    \begin{equation}
        \tau > F_\mathrm{adi}\tau_0,
        \label{Eq_adiabatic_condition_CMAT}
    \end{equation}
    where $\tau_0$ is defined as
    \begin{equation} 
        \tau_0 =
        \text{max}
        \left\{
            \frac{
                g{\Omega_0}^2\left| A_1(t) - B_1(t)e^{2t/\tau}  \right|e^{t/\tau}
            }{
                \left|\omega_{\pm1}(t)\right|\sqrt{N_0(t)N_1(t)}\left(1+e^{2t/\tau}\right)^2
            }
        \right\}
        \label{Eq_tau0}
    \end{equation}
    (see Appendix~\ref{appendix_Adiabatic-Condition} for the detailed derivation).
    The coefficient $F_\mathrm{adi}$ is an adiabaticity factor, $N_0(t)$ and $N_1(t)$ are normalization factors as shown in Eqs.~\eqref{appendix_Eq_phi0}~and~\eqref{appendix_Eq_phi1}, 
    and $A_1(t)$ and $B_1(t)$ are defined by Eq.~\eqref{appendix_Eq_A1B1A2B2} in Appendix~\ref{appendix_eigen_derivation}.
    On the other hand, the condition for suppressing the population of the cavity field, $\Omega_0/g\ll1$, can be read as a condition of $\tau$ via a relation between $\Omega_0$ and $\tau$, which appears by optimizing the operation of the control laser to maximize the success probability of the CMAT.

    We analytically determine which of these two conditions more severely limits the operational speed.
    The adiabatic condition represented by Eqs.~\eqref{Eq_adiabatic_condition_CMAT} and \eqref{Eq_tau0} is rewritten in a simpler representation by assuming $C\gg1$, $\tau > F_\mathrm{adi}\tau_0$ and $\Omega_0/g\ll1$~\cite{Goto2008}. With these assumptions, the adiabatic condition reduces to
    \begin{equation}
        \tau > F_\mathrm{adi}\tau_0,\ \ \ \ 
        \tau_0 \approx \frac{1}{\Omega_0}\cdot\mathrm{max}\left\{
            \frac{e^{t/\tau}}{1+e^{2t/\tau}}
        \right\}
        = \frac{1}{2\Omega_0}.
        \label{Eq_T0_Lcav_short}
    \end{equation}
    By substituting the balancing condition Eq.~\eqref{Eq_PL-balancing-condition} to Eq.~\eqref{Eq_T0_Lcav_short}, the adiabatic condition is written with cavity parameters as
    \begin{equation}
        \tau > \widetilde{\tau}_\mathrm{a},\ \ \ \ 
        \widetilde{\tau}_\mathrm{a}={F_\mathrm{adi}}^2 \cdot \frac{1}{4g}\sqrt{\frac{\kappa}{2\gamma}}=\frac{{F_\mathrm{adi}}^2}{8\gamma\sqrt{C}}.
        \label{Eq_Fadi-adiabatic-condition_PB}
    \end{equation}
    The condition for suppressing the population of cavity field, $\Omega_0/g\ll 1$, is rewritten by substituting the balancing condition as
    \begin{equation}
        \tau > \widetilde{\tau}_\mathrm{c},\ \ \ \ \ \ 
        \widetilde{\tau}_\mathrm{c} = \frac{1}{{F_\mathrm{cp}}^2} \cdot \frac{1}{g}\sqrt{\frac{2\gamma}{\kappa}}
        =\frac{1}{{F_\mathrm{cp}}^2}\cdot
        \frac{2\gamma\sqrt{C}}{g^2},
        \label{Eq_Fcp}
    \end{equation}
    where we have introduced a parameter $F_\mathrm{cp}$ denoting the degree of suppression of cavity-field population, that is, $\Omega_0/g\ll 1$ is rewritten as $\Omega_0/(gF_\mathrm{cp}) < 1,\ F_\mathrm{cp} \ll 1$.
    Since the modified adiabatic condition \eqref{Eq_Fadi-adiabatic-condition_PB} assumes $\Omega_0/g\ll1$, it is only valid in the CQED-parameter region where the adiabatic condition is more strongly limiting the operational speed than the condition for suppressing the population of the cavity field. 
   Despite this, the conditions in Eqs.~\eqref{Eq_Fadi-adiabatic-condition_PB} and \eqref{Eq_Fcp} at least allow us to determine the CQED-parameter region where the adiabatic condition is more severe.

\begin{figure}[b]
        \centering
        \includegraphics[clip,width=8cm]{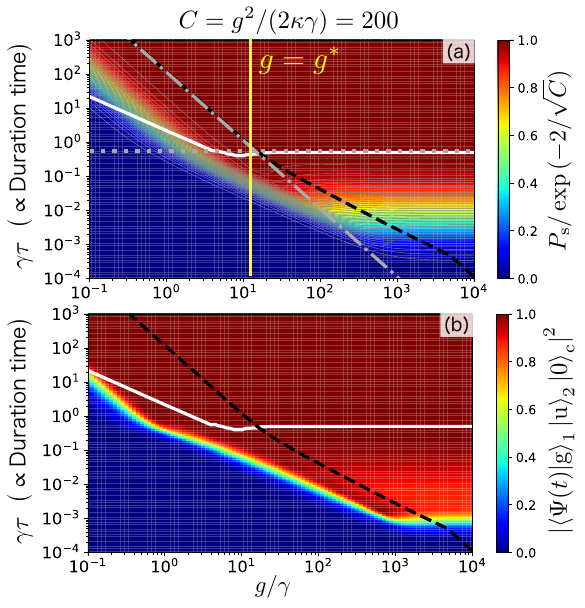}
        \caption{(a) $\tau=\widetilde{\tau}_\mathrm{a}$ in Eq.~\eqref{Eq_Fadi-adiabatic-condition_PB} $\tau=\widetilde{\tau}_\mathrm{c}$ in Eq.~\eqref{Eq_Fcp} for $F_\mathrm{adi}=8$ and $F_\mathrm{cp}=0.5$ represented by the gray dotted line and the gray dashed-dotted line, respectively. Color graph represents the success probability normalized by the upper bound $\exp({-2/\sqrt{C}})$. (b) Overlap with the ideal final state at the end of the process.
        The black dashed curve represents the line that satisfies $\Omega_0/g=0.5$. The white solid curve represents $\tau=F_\mathrm{adi}\tau_0$, where $F_\mathrm{adi}=8$.
        For the numerical calculation, we truncate the operational period from $t=-7.5\tau$ to $7.5\tau$ according to Appendix~\ref{appendix_normalized_Rabi_frequency}. We fix $C=200$, then the upper bound $\exp({-2/\sqrt{C}})\approx0.868$, and the maximal Rabi frequency $\Omega_0$ is optimized to maximize the success probability.
        }
        \label{Main}
    \end{figure}

    By comparing Eq.~\eqref{Eq_Fadi-adiabatic-condition_PB} and Eq.~\eqref{Eq_Fcp}, the adiabatic condition is the more limiting condition when the amplitude decay rate $\kappa$ is larger than
    \begin{equation}
        \kappa^* \coloneqq \frac{8\gamma}{\left(F_\mathrm{adi}F_\mathrm{cp}\right)^2}.
        \label{Eq_opt-kappa}
    \end{equation}
    In the case of the upper bound of the success probability being fixed, namely, for a fixed $C$, the above discussion is also provided according to the coupling constant $g$; the adiabatic condition is the more severe condition when the coupling constant $g$ is larger than
    \begin{equation} 
        g^* \coloneqq \frac{4\gamma\sqrt{C}}{F_\mathrm{adi}F_\mathrm{cp}}.
        \label{Eq_opt-g}
    \end{equation}
    In Fig.~\ref{Main} (a), we plot the lines $\tau=\widetilde{\tau}_\mathrm{c}$ and $\tau=\widetilde{\tau}_\mathrm{a}$ as a function of $g/\gamma$ with $C$ fixed at $200$, which clearly shows that the adiabatic condition is more severe than the condition for suppressing the population of the cavity field when $g>g^*$.

    \subsection{Numerical simulations\label{sec_numerical-results}}
    In Fig.~\ref{Main}, we plot the success probability normalized by the upper bound $\exp(-2/\sqrt{C})$ solving the Schr\"{o}dinger-like equation in Eq.~\eqref{Eq_TimeEvolutionEquation} following the effective Hamiltonian~\eqref{Eq_effectiveHamiltonian}. The success probability is calculated from $(1-I_\mathrm{a}-I_\mathrm{cav})\left|\braket{\Psi(t=\infty)|\mathrm{g}}_1\Ket{\mathrm{u}}_2\Ket{0}_\mathrm{c}\right|^2$ at the end of the process, where $I_\mathrm{a}$ and $I_\mathrm{cav}$ are defined by
    \begin{eqnarray}
        I_\mathrm{a} &=& 2\gamma\int dt \biggl(\left|\braket{\widetilde{\Psi}(t)|\mathrm{e}}_1\Ket{\mathrm{g}}_2\Ket{0}_\mathrm{c}\right|^2\nonumber\\
        &&\ \ \ \ \ \ \ \ \ \ \ \ \ \ + \left|\braket{\widetilde{\Psi}(t)|\mathrm{g}}_1\Ket{\mathrm{e}}_2\Ket{0}_\mathrm{c}\right|^2 \biggr),\\
        I_\mathrm{cav} &=& 2\kappa\int dt\left|\braket{\widetilde{\Psi}(t)|\mathrm{g}}_1\Ket{\mathrm{g}}_2\Ket{1}_\mathrm{c}\right|^2.
    \end{eqnarray}
    The maximum Rabi frequency $\Omega_0$ is optimized to maximize the success probability, and the cooperativity is fixed as $C=200$ to fix the upper bound $\exp({-2/\sqrt{C}})$.
    The white solid curve represents $\tau=F_\mathrm{adi}\tau_0$ with the definition of $\tau_0$ in Eq.~\eqref{Eq_tau0}, and if the operational speed is faster, namely, $\tau$ is shorter, than this threshold, the population of the excited state increases due to the violation of the adiabatic condition. The black dashed curve represents $\Omega_0/g=F_\mathrm{cp}$, and if the operational speed is faster than this threshold, the population of cavity photons increases. 
    These curves are calculated numerically without any approximations.
    The equation $\Omega_0/g=F_\mathrm{cp}$ is effectively transformed into the equation of $\tau$ through the optimization of the operation of the control laser to maximize the success probability.
    
    It is observed in Fig.~\ref{Main}(a) that the success probability deviates significantly from the upper bound when the operation of the control laser is performed faster than at least one of the thresholds. For the parameter set in Fig.~\ref{Main}(a), namely, $F_\mathrm{cp}=0.5$ and $F_\mathrm{adi}=8$, the success probability is over $0.99\exp({-2/\sqrt{C}})$ when $\tau$ is longer than both of the thresholds.
    As the analytical results indicate, the condition for suppressing the population of the cavity field is more strongly limiting the operational speed compared to the adiabatic condition for $g<g^*$, while the adiabatic condition is more severe for $g>g^*$.
    The maximal operational speed is achieved for $g>g^*$, whose value is $\widetilde{\tau}_\mathrm{a}=\frac{{F_\mathrm{adi}}^2}{8\gamma\sqrt{C}}$. To achieve large $g$ for fixed $C$, for example, we can reduce cavity length~\cite{Asaoka2021, Utsugi2022,Utsugi2022arxiv}.
    The discrepancy between the analytical (gray lines) and numerical (white and black curves) results stems from the violation of the approximations in the analysis such that the balancing condition used in the analysis does not maximize the success probability except for the region where the success probability is close to the upper bound.
    
    Figure~\ref{Main}(b) plots the fidelity of CMAT, namely $\left|\braket{\Psi(t=\infty)|\mathrm{g}}_1\Ket{\mathrm{u}}_2\Ket{0}_\mathrm{c}\right|^2$. This conditional success probability appears, for instance, when photon loss events are eliminated with photon detection~\cite{Goto2010}.
    In this case, the region with success probability close to the upper bound expands compared to the case without photon detection, that is, Fig~\ref{Main}(a). This indicates that photon detection enables a faster operation in CMAT.

\section{Conclusion\label{sec_Conclusion}}
    
    In this paper, we have investigated the operational speed limit of CMAT conditioned on the success probability being close to the upper bound.
    The operational speed is limited by the two main factors, formulated as $\widetilde{\tau}_\mathrm{c}$ and $\widetilde{\tau}_\mathrm{a}$, where $\widetilde{\tau}_\mathrm{a}$ is attributed to the adiabatic condition and $\widetilde{\tau}_\mathrm{c}$ is attributed to the condition for suppressing the cavity decay.
    We have shown that the dominant factor limiting the operational speed is $\widetilde{\tau}_\mathrm{c}$ for $\kappa<\kappa^*$, while $\widetilde{\tau}_\mathrm{a}$ for $\kappa>\kappa^*$.
    The maximal operational speed of CMAT is achieved for $\kappa>\kappa^*$, being proportional to $1/\widetilde{\tau}_\mathrm{a}$, namely, $\gamma\sqrt{C}$, when the cooperativity $C$ is fixed.

\section*{Acknowledgement}
The authors thank Hayato Goto, Akihisa Goban, Samuel Ruddell, Karen Webb, Donald White, and Seigo Kikura for their helpful comments.
This work is supported by JST CREST, Grant Number JPMJCR1771, JST Moonshot R\&D, Grant Numbers JPMJMS2061, JPMJMS2268, and JST SPRING, Grant Number JPMJSP2128, Japan.

\appendix
\begin{widetext}

    \section{Derivation of eigenenergies and eigenstates in CMAT\label{appendix_eigen_derivation}}
        In CMAT, since the initial state of CMAT is in $\Ket{\mathrm{u}}_1\Ket{\mathrm{g}}_2\Ket{0}_\mathrm{c}$, the time evolution of the system can be described with the subspace spanned by the five basis states: \{ $\Ket{\mathrm{u}}_1\Ket{\mathrm{g}}_2\Ket{0}_\mathrm{c}$, $\Ket{\mathrm{g}}_1\Ket{\mathrm{u}}_2\Ket{0}_\mathrm{c}$, 
    $\Ket{\mathrm{e}}_1\Ket{\mathrm{g}}_2\Ket{0}_\mathrm{c}$, 
    $\Ket{\mathrm{g}}_1\Ket{\mathrm{e}}_2\Ket{0}_\mathrm{c}$, 
    $\Ket{\mathrm{g}}_1\Ket{\mathrm{g}}_2\Ket{1}_\mathrm{c}$ \}.
    Under this set of states, the Hamiltonian given by Eq.~\eqref{Hamiltonian} is expressed in matrix form as 
    \begin{equation}
        \mathcal{H}(t) = \mathrm{i} \hbar
        \left(
        \begin{array}{ccccc}
        0 & 0 & -\Omega_1(t) & 0 & 0  \\
        0 & 0 & 0 & -\Omega_2(t) & 0  \\
        \Omega_1(t) & 0 & 0 & 0 & g  \\
        0 & \Omega_2(t) & 0 & 0 & g  \\
        0 & 0 & -g & -g & 0 
        \end{array}
        \right).
        \label{Hamiltonian_matrixForm}
    \end{equation}
        Therefore, the eigenenergies and eigenstates of the Hamiltonian in Eq.~\eqref{Hamiltonian_matrixForm} are given by
        \begin{eqnarray}
            \omega_0 &=& 0,\\
            \Ket{\phi_0(t)} &=& \Ket{\mathrm{D}(t)}=
            \frac{1}{\sqrt{N_0(t)}}
    		\left(
    		\begin{array}{c}
    		g\Omega_2(t)\\
    		g\Omega_1(t)\\
    		0\\
    		0\\
    		- \Omega_1(t)\Omega_2(t)
    		\end{array}
    		\right),\label{appendix_Eq_phi0}\\
    		\omega_{\pm1}(t) &=& \pm \sqrt{\frac{2g^2+{\Omega_1(t)}^2+{\Omega_2(t)}^2-\sqrt{4g^4+\left({\Omega_1(t)}^2-{\Omega_2(t)}^2\right)^2}}{2}},\ \ \ \ \ \\
    		\Ket{\phi_{\pm1}(t)} &=&
    		\frac{1}{\sqrt{N_1(t)}}
    		\left(
    		\begin{array}{c}
    		A_1(t)\Omega_1(t)\\
    		B_1(t)\Omega_2(t)\\
    		\mathrm{i}\omega_{\pm1}(t)A_1(t)\\
    		\mathrm{i}\omega_{\pm1}(t)B_1(t)\\
    		g\bigl(A_1(t)+B_1(t)\bigr)
    		\end{array}
    		\right),\label{appendix_Eq_phi1}\\
    		\omega_{\pm2}(t) &=& \pm \sqrt{\frac{2g^2+{\Omega_1(t)}^2+{\Omega_2(t)}^2+\sqrt{4g^4+\left({\Omega_1(t)}^2-{\Omega_2(t)}^2\right)^2}}{2}},\\
    		\Ket{\phi_{\pm2}(t)} &=&
    		\frac{1}{\sqrt{N_2(t)}}
    		\left(
    		\begin{array}{c}
    		A_2(t)\Omega_1(t)\\
    		B_2(t)\Omega_2(t)\\
    		\mathrm{i}\omega_{\pm2}(t)A_2(t)\\
    		\mathrm{i}\omega_{\pm2}(t)B_2(t)\\
    		g\bigl(A_2(t)+B_2(t)\bigr)
    		\end{array}
    		\right)\label{appendix_Eq_phi2},
        \end{eqnarray}
        where $N_k(t)\ (k=0,1,2)$ indicates normalization factor, and $A_1(t)$, $B_1(t)$, $A_2(t)$ and $B_2(t)$ are defined as
        \begin{equation}
            \begin{array}{ccc}
            A_1(t) & = & {\Omega_2(t)}^2 -{\omega_{\pm1}(t)}^2 + 2g^2, \\
            B_1(t) & = & - {\Omega_1(t)}^2 + {\omega_{\pm1}(t)}^2 - 2g^2, \\
            A_2(t) & = & {\Omega_2(t)}^2 - {\omega_{\pm2}(t)}^2, \\
            B_2(t) & = & {\Omega_1(t)}^2 - {\omega_{\pm2}(t)}^2.
            \end{array}
        \label{appendix_Eq_A1B1A2B2}
        \end{equation}
        These expressions are also used in Ref.~\cite{Goto2008}.
        
        Here we derive the above eigenstates.
        Note that the time-dependent notation of each parameter is omitted below for simplicity.
        We assume that $\bm{x}=(x_1, x_2, x_3, x_4, x_5)^\top$ is an eigenstate of the Hamiltonian in Eq.~\eqref{Hamiltonian_matrixForm}.
        Since $\hbar\omega_j(j=0,\pm1,\pm2)$ are the eigenenergies of the Hamiltonian, $\bm{x}$ satisfies $\mathcal{H}(t)\bm{x}=\omega_j\bm{x}$. 
        Therefore, we have
        \begin{empheq}[left={\empheqlbrace}]{alignat=2}
            \mathrm{i} \omega_j x_1 & \quad = \Omega_1 x_3\label{appendix_eqq_x_1}\\
            \mathrm{i} \omega_j x_2 & \quad = \Omega_2 x_4\label{appendix_eqq_x_2}\\
            \mathrm{i} \omega_j x_3 & \quad = - \Omega_1 x_1 - g x_5\label{appendix_eqq_x_3}\\
            \mathrm{i} \omega_j x_4 & \quad = - \Omega_2 x_2 - g x_5\label{appendix_eqq_x_4}\\
            \mathrm{i} \omega_j x_5 & \quad = g (x_3 + x_4)
            \label{appendix_eqq_x_5}.
        \end{empheq}
        
        In case of $j=0$, i.e., $\omega_j = 0$, $\Omega_1 x_1 = -g x_5$, $\Omega_2 x_2 = -g x_5$ and $x_3=x_4=0$ are required. 
        Thus, the eigenstate corresponding to $\omega_0$ is proportional to $\bm{x}=(g\Omega_2, g\Omega_1, 0, 0, -\Omega_1\Omega_2)^\top$.
        This is the dark state given by~\eqref{Eq_darkState}.
        
        On the other hand, in case of $j\neq0$ ($\omega_j \neq 0$), the eigenstate corresponding to $\omega_j$ is proportional to 
        \begin{equation}
            \left(
            \begin{array}{c}
            \Omega_1 x_3\\
            \Omega_2 x_4\\
            \mathrm{i} \omega_j x_3\\
            \mathrm{i} \omega_j x_4\\
            g (x_3 + x_4)
            \end{array}
            \right).
        \end{equation}
        Note that $x_3$ and $x_4$ are not independent because they are restricted by
        \begin{empheq}[left={\empheqlbrace}]{alignat=2}
        \left( {\Omega_1}^2 + g^2 -{\omega_j}^2 \right) x_3 & \quad =\ \ \  g^2 x_4\label{appendix_eqq_x3x4} \\
        \left( {\Omega_2}^2 + g^2 -{\omega_j}^2 \right) x_4 & \quad =\ \ \  g^2 x_3.\label{appendix_eqq_x4x3}
        \end{empheq}
        These relations can be derived by substituting $x_1$, $x_2$ and $x_5$, obtained from Eq.~\eqref{appendix_eqq_x_1}, \eqref{appendix_eqq_x_2} and~\eqref{appendix_eqq_x_5}, into Eq.~\eqref{appendix_eqq_x_3} and~\eqref{appendix_eqq_x_4}.
        By calculating~\eqref{appendix_eqq_x3x4} $+$ \eqref{appendix_eqq_x4x3}, we have
        \begin{equation}
            \left(
                {\Omega_2}^2 - {\omega_j}^2 +2g^2
            \right)x_4
            =
            - \left(
                {\Omega_1}^2 - {\omega_j}^2 +2g^2
            \right)x_3.
        \end{equation}
        This relation implies
        \begin{equation}
            \left(
            \begin{array}{c}
                x_3\\
                x_4
            \end{array}
            \right)
            \propto
            \left(
            \begin{array}{c}
                {\Omega_2}^2 - {\omega_j}^2 +2g^2\\
                -{\Omega_1}^2 + {\omega_j}^2 -2g^2
            \end{array}
            \right),
        \end{equation}
        and corresponds to $A_1$ and $B_1$ in Eq.~\eqref{appendix_Eq_A1B1A2B2} when $\omega_j=\omega_{\pm1}$.
        Note that this relation can not be applied when $\omega_j=\omega_{\pm2}$ because ${\Omega_2}^2 - {\omega_{\pm2}}^2 +2g^2$ and ${\Omega_1}^2 - {\omega_{\pm2}}^2 +2g^2$ become zero when $\Omega_1=\Omega_2$.
        Subsequently, by calculating~\eqref{appendix_eqq_x3x4} $-$ \eqref{appendix_eqq_x4x3}, we also have
        \begin{equation}
            \left(
                {\Omega_2}^2 - {\omega_j}^2
            \right)x_4
            =
            \left(
                {\Omega_1}^2 - {\omega_j}^2
            \right)x_3.
        \end{equation}
        This relation implies
        \begin{equation}
            \left(
            \begin{array}{c}
                x_3\\
                x_4
            \end{array}
            \right)
            \propto
            \left(
            \begin{array}{c}
                {\Omega_2}^2 - {\omega_j}^2\\
                {\Omega_1}^2 - {\omega_j}^2
            \end{array}
            \right),
        \end{equation}
        and corresponds to $A_2$ and $B_2$ in Eq.~\eqref{appendix_Eq_A1B1A2B2} when $\omega_j=\omega_{\pm2}$.
        Note that this relation can not be applied when $\omega_j=\omega_{\pm1}$ because ${\Omega_2}^2 - {\omega_{\pm1}}^2$ and ${\Omega_1}^2 - {\omega_{\pm1}}^2$ become zero when $\Omega_1=\Omega_2$.
 
    \section{Adiabatic condition
    \label{appendix_Adiabatic-Condition}}

    Here, we derive the adiabatic condition for CMAT  in the absence of photon loss for simplicity.
    We apply CMAT to the common adiabatic condition
    \begin{equation}
        \left|
            \Braket{m(t)|\frac{d\ }{dt}|n(t)}
        \right|
        \ll
        \frac{1}{\hbar} \left| E_m(t) - E_n(t)\right|
        ,
        \label{Eq_generalAdiabaticCondition}
    \end{equation}
    which is given by Ref.~\cite{MessiahBook1978}.
    Here, $E_{n}(t)$ and $E_{m}(t)$ are the eigenenergies of the Hamiltonian at time $t$ ($n\neq m$), and $\Ket{n(t)}$ and $\Ket{m(t)}$ are the eigenstates of the Hamiltonian corresponding to $E_{n}(t)$ and $E_{m}(t)$, respectively.
    According to the adiabatic theorem, an initially prepared state $\Ket{n(0)}$ remains in state $\Ket{n(t)}$ under Hamiltonian evolution, and the probability to deviate from $\Ket{n(t)}$ to $\Ket{m(t)}$ can be suppressed when the condition given by Eq.~\eqref{Eq_generalAdiabaticCondition} is satisfied.
    
    By substituting $\hbar\omega_{0,\pm1}$ and $\Ket{\phi_{0,\pm1}}$ to the condition in Eq.~\eqref{Eq_generalAdiabaticCondition}, we obtain the adiabatic condition to keep the state in $\Ket{\mathrm{D}(t)}$:
    \begin{equation}
        \frac{g{\Omega_0}^2\left| A_1(t) \dot{f}_1(t) f_2(t) + B_1(t)f_1(t)\dot{f}_2(t) \right|}
        {\sqrt{N_0(t)N_1(t)}}
        \ll
        \left|\omega_{\pm1}(t)\right|.
        \label{Eq_adiabatic-condition-for-CMAT}
    \end{equation}
    Note that as the probability to deviate from $\Ket{\mathrm{D}}$ to $\Ket{\phi_{\pm2}}$ is less than that to $\Ket{\phi_{\pm1}}$ due to $\left|\omega_{\pm1}(s)\right|<\left|\omega_{\pm2}(s)\right|$, we only focus on the case of $j=\pm1$.
    By applying the typical functions in Eq.~\eqref{Eq_concrete_f1f2}, the adiabatic condition in Eq.~\eqref{Eq_adiabatic-condition-for-CMAT} is rewritten as
    \begin{equation}
        \tau \gg \tau_0, 
        \label{App_Eq_adiabatic_condition_CMAT}
    \end{equation}
    where $\tau_0$ is defined as
    \begin{equation} 
        \tau_0 =
        \text{max}
        \left\{
            \frac{
                g{\Omega_0}^2\left| A_1(t) - B_1(t)e^{2t/\tau}  \right|e^{t/\tau}
            }{
                \left|\omega_{\pm1}(t)\right|\sqrt{N_0(t)N_1(t)}\left(1+e^{2t/\tau}\right)^2
            }
        \right\}.
        \label{App_Eq_tau0}
    \end{equation}
    Note that the right side of Eq.~\eqref{App_Eq_tau0} is not influenced by $\tau$ since only the maximum value from $t=-\infty$ to $\infty$ is taken.
    
    As shown in Eq.~\eqref{App_Eq_adiabatic_condition_CMAT}, $\tau_0$ describes the limit on the operation time for performing CMAT adiabatically.
    The validity of this limit can be observed in numerical simulations.
    Figure~\ref{CMAT_withoutDissipation} represents the numerically calculated success probability by solving the Schr\"{o}dinger-like equation in Eq.~\eqref{Eq_TimeEvolutionEquation}, where $\gamma=\kappa=0$.
    According to the result in Fig.~\ref{CMAT_withoutDissipation}, the adiabatic condition of CMAT can be modified as
    \begin{equation}
        \tau > F_\mathrm{adi}\ \tau_0,
        \label{Eq_modified_adiabatic_condition_CMAT}
    \end{equation}
    where $F_\mathrm{adi} \approx 8$ (white line) is the adiabaticity factor, such that a fidelity over $0.99$ is achieved.
    It is also seen from Fig.~\ref{CMAT_withoutDissipation} that the adiabatic condition is approximately given by $\Omega_0 \tau, g\tau\gg1$, consistent with the past observations discussed in Refs.~\cite{Pellizzari1995, Sangouard2005, Sangouard2006}.
    Therefore, $\tau_0$ can be reduced, which is equivalent to relaxing the adiabatic condition, by increasing both the maximum Rabi frequency $\Omega_0$ and the coupling constant $g$.

    \begin{figure}[t]
        \centering
        \includegraphics[clip,width=8cm]{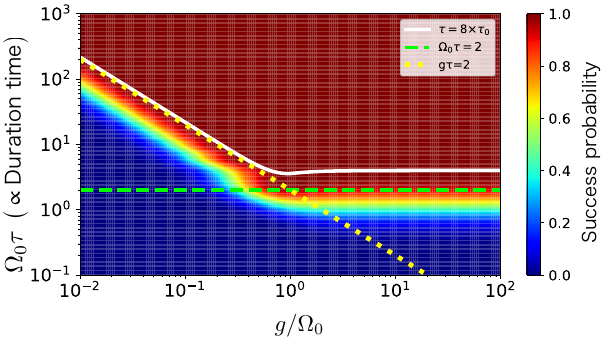}
        \caption{Fidelity of CMAT in the absence of dissipation, obtained by numerically solving the Schr\"{o}dinger-like equation in Eq.~\eqref{Eq_TimeEvolutionEquation}, where $\gamma=\kappa=0$.
        The green dashed and yellow dotted lines represent lines that satisfy $\Omega_0 \tau = 2$ and $g \tau = 2$, respectively.
        The white solid curve represents $\tau=F_\mathrm{adi}\ \tau_0$, where $F_\mathrm{adi}=8$.
        }
        \label{CMAT_withoutDissipation}
    \end{figure}

    \section{Truncation of CMAT process \label{appendix_normalized_Rabi_frequency}}
    For the numerical simulations in this paper, we truncate the period of CMAT from $t=-7.5\tau$ to $7.5\tau$.
    The truncated period corresponds to the actual duration time $T$, as $T=15\tau$ in Fig.~\ref{AppFig_Rabi-frequency_shape} (a).
    As shown in Fig.~\ref{AppFig_Rabi-frequency_shape} (b), the fidelity of CMAT is close to 1 by increasing the actual duration time $T$ compare to the time constant $\tau$, since the shorter $T$ is less adiabatic.
    We use $T=15\tau$ as a truncation period long enough to implement CMAT adiabatically.
    
    \begin{figure}[htbp]
        \centering
        \includegraphics[clip,width=14cm]{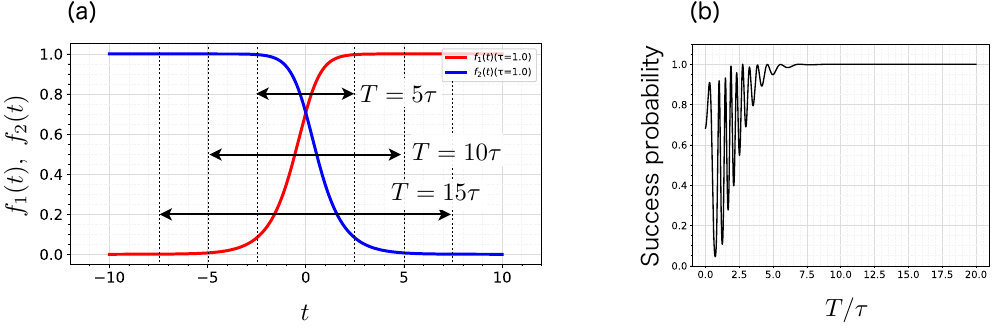}
        \caption{(a) The shape of $f_1(t)$ and $f_2(t)$, where $\tau=1$.
        (b) The fidelity of CMAT depending on $T/\tau$. 
        This result is obtained by the numerical calculation under the condition of $g=\Omega_0$ and $\Omega_0 T = 1000$, where the dissipation is not taken into account.
        }
        \label{AppFig_Rabi-frequency_shape}
    \end{figure}. 

\end{widetext}

\bibliography{suenaga_paper}

\end{document}